# Cost & Capability Compromises in STEM Instrumentation for Low-Voltage Imaging

*Frances Quigley[1,2*], Patrick McBean[1,2], Peter O'Donovan[1], Jonathan J. P. Peters[1,2], Lewys Jones[1,2]*

1. School of Physics, Trinity College Dublin, Dublin 2, Ireland

2. Advanced Microscopy Laboratory, Centre for Research on Adaptive Nanostructures & Nanodevices (CRANN), Dublin 2, Ireland

## Abstract

Low voltage transmission electron microscopy ($\leq 80$ kV) has many applications in imaging beam-sensitive samples, such as metallic nanoparticles, which may become damaged at higher voltages. To improve resolution, spherical aberration can be corrected for in a Scanning Transmission Electron Microscope (STEM), however chromatic aberration may then dominate, limiting the ultimate resolution of the microscope. Using image simulations, we examine how a chromatic aberration corrector, different objective lenses, and different beam energy-spreads each affect the image quality of a gold nanoparticle imaged at low voltages in a spherical aberration-corrected STEM. Quantitative analysis of the simulated examples can inform the choice of instrumentation for low-voltage imaging. We here demonstrate a methodology whereby the optimum energy spread to operate a specific STEM can be deduced. This methodology can then be adapted to the specific sample and instrument of the reader, enabling them to make an informed economical choice as to what would be most beneficial for their STEM in the cost-conscious landscape of scientific infrastructure.

**Keywords:** scanning transmission electron microscope (STEM), monochromation, low-voltage imaging, chromatic aberration, image simulation

*Corresponding Author: fquigley@tcd.ie

## Introduction

From medical imaging to beamlines or electron microscopy, any time radiation is used to probe matter we risk the chance of unwanted transformations. In electron microscopy we refer to these unwanted transformations as beam damage. For example, in the scanning transmission electron microscope (STEM) many samples, such as metallic nanoparticles, are prone to knock-on damage when imaged at high voltages (Egerton, 2019; Azcárate et al., 2017). Metallic nanoparticles are of widespread interest to many different areas of research due to their optical (Jain et al., 2008) and catalytic (Louis & Pluchery, 2012) properties, and use in biomedical sciences such as for bioimaging (Sharma et al., 2006) or drug delivery (Dreaden et al., 2012), so approaches to minimise beam damage have been keenly investigated (Van Aert et al., 2019). Cooling certain samples has been shown to decrease knock-on damage (Egerton, 2019). However, cryogenic TEM can be expensive (Kuntsche et al., 2011), and cryogenic sample holders are often not compatible with other stimuli such as biasing, so these are not the focus of this manuscript. We will instead be concentrating on the effect of lowering the electron acceleration voltage, which is already a well-known technique to reduce the effect of knock-on damage, as well as providing higher scattering contrast for thin specimens (Egerton, 2014).

The effects of lowering the acceleration voltage also manifest themselves because the STEM is not a perfect optical system. Its image resolution depends on the entire lens system - including the cumulative effect of spherical aberration ($C_s$) and chromatic aberration ($C_c$) (Klie, 2009), as well as the diffraction limit (Scherzer, 1949). Figure 1 shows the aberration and diffraction-limited contributions to probe size as a function of probe semi-angle (a variety of condenser-apertures are available to a STEM operator), and the resultant expected resolution. For a system constrained by spherical aberration the optimum operating condition is shown by Point 1. While $C_s$ was once the main limitation to beam diameter, spherical aberration correctors (Krivanek et al., 1997; Haider et al., 1998) have become increasingly common in STEMs. $C_s$ correctors operate by reducing the $C_3$ coefficient by an arrangement of non-rotationally symmetric lenses (Scherzer, 1949).

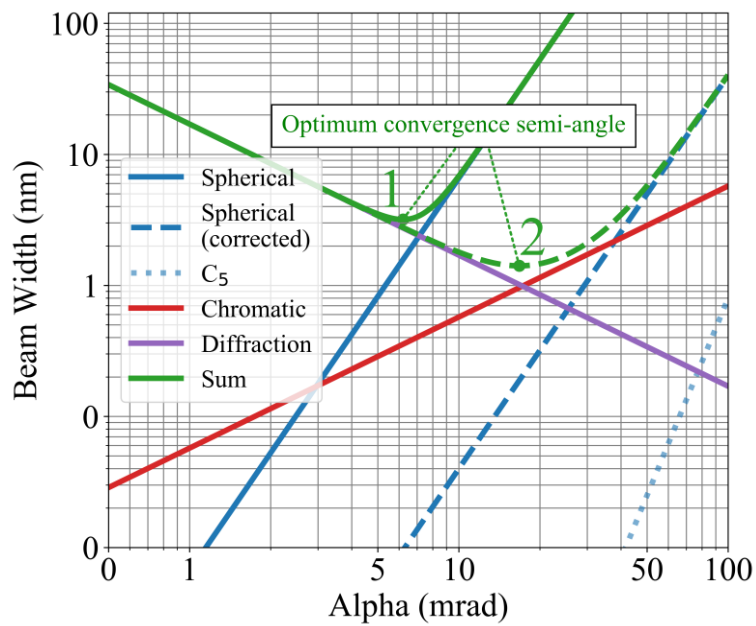

*Figure 1: Contributions to beam diameter as a function of probe semi-angle from spherical and chromatic aberrations, and the diffraction limit. In this example, the energy spread of the microscope is 287 meV and its acceleration voltage is 30 kV. For point 1, the $C_s$ coefficient is 3.3 mm, while for point 2, the $C_c$ coefficient is 3.0 mm.*

With $C_3$ corrected, $C_5$, a higher order spherical aberration, can become the next limitation, or at lower voltages chromatic aberration dominates. Point 2 in Figure 1 represents a $C_s$-corrected STEM where the $C_3$ coefficient was reduced from 3.0 mm to 0.02 mm which subsequently causes chromatic aberration to become the limiting factor to STEM probe size. If the effects of chromatic aberration are reduced, $C_5$ would limit the ultimate resolution of the system (Schramm et al., 2012), however at that point the resolution would already be acceptable for low voltage imaging.

Chromatic aberration occurs when lower energy electrons are focused prematurely to the optic axis relative to higher energy electrons (Klie, 2009). It can be defined in terms of the diameter of the blurred disk of electrons formed $d_{chr}$;

$$d_{chr} = C_c \frac{\Delta E}{E} \alpha \qquad (1)$$

where $C_c$ is the chromatic aberration coefficient, $\alpha$ is the beam convergence semi-angle, and $\Delta E$ and $E$ are the electron energy spread and electron energy respectively (Goodhew et al., 2000). Due to its inverse dependence on acceleration voltage, chromatic aberration becomes increasingly significant at lower voltages. Reducing chromatic defocus blur is therefore extremely advantageous for high resolution low voltage imaging in the $C_s$-corrected STEM. There are three main options to minimise the effects of chromatic aberration which, importantly, we will discuss in the context of their economic feasibility and practicality of operation.

The first and perhaps most obvious is to correct the $C_c$ by installing a chromatic aberration corrector into the STEM. From a simulation perspective, chromatic aberration correctors are numerically the same as using a STEM lens with a much smaller inherent $C_c$ coefficient. Therefore, an objective lens with a $C_c$ coefficient of 0 mm (Linck et al., 2016) has been included in our following simulations. Chromatic aberration correctors however require a significant amount of extra training for the user and are currently predominantly developmental projects which by their very nature means they are extremely expensive. These are important factors to consider in their comparison with the other hardware solutions available.

The second option for reducing the chromatic defocus blur is decreasing the $C_c$ coefficient of the objective lens itself. The resolution in a STEM is limited by the objective lens (OL) as the $C_s$ and $C_c$ coefficients increase significantly with the pole piece gap of the lens (Tsuno & Jefferson, 1998). During the tendering process of a STEM, several key specifications of the microscope must be decided upon, including the pole piece gap of the OL. Therefore, by purchasing a STEM with an OL configuration with a smaller $C_c$ coefficient, one can reduce the effect of chromatic aberration. While higher resolution can be attained by using an OL with a smaller pole piece gap, it comes at the cost of flexibility; reduced sample tilt range, lower EDX collection efficiency and the exclusion of the use of certain specialist holders. Therefore, the samples which will be analysed in the STEM must be considered when deciding on the OL configuration to purchase. However, since this decision is made during the tendering process, this option in reducing the $C_c$ coefficient is not available unless you are in the process of purchasing a new STEM.

Lastly, one can reduce the chromatic defocus blur by reducing the energy spread of the electrons. This can be achieved by replacing an electron source with a large energy spread, such as a thermionic tungsten electron gun (~3,000 meV) (Carter & Williams, 2016), with a source with a lower electron energy spread such as a cold Field Emission Gun (cold FEG) (~200-400 meV) (Kimoto & Matsui, 2002). Thermionic electron sources tend not to be equipped in high end instruments which include spherical aberration correctors, so a Schottky

FEG (600-800 meV) (Naydenova et al., 2019) will be the largest energy spread electron source that we will investigate, along with a cold FEG for comparison. Cold FEGs are brighter than Schottky FEGs by a factor of 1.1-2.0 (Xin et al., 2013; Carter & Williams, 2016) so a mid-range value of 1.5 was used for simulating the difference in brightness between these two FEGs. A monochromator could also be installed in the system if an even smaller electron energy spread is required. Unfortunately, these reject the more energy-dispersed electrons, reducing the beam current and ultimately may degrade the signal-to-noise ratio (SNR) (Hachtel et al., 2018). While scanning slower is an option to compensate for this loss of SNR, this may become counterproductive with the resulting increase in environmental instabilities which can degrade the image (Muller & Grazul, 2001; Muller et al., 2006). A balance therefore must be struck between electron dose and energy spread requirements.

The aim of this research is to determine a reasonable compromise solution to reducing the effects of chromatic aberration in a $C_s$-corrected microscope while also considering the financial cost of the infrastructure. We will explore this by first determining the beam current remaining after increasing levels of monochromation for both a cold and Schottky FEG and then simulating images representing different trade-offs of dose and chromatic spread. These images are then quantitatively evaluated using their signal-to-noise ratios before some general recommendations are made. The aim of this work is not to speculate what a perfect instrument bought with unlimited equipment budget might achieve, but to guide what might be practically achievable for the majority of research centres. This is therefore a methodology which the reader can follow to draw a conclusion for their own specific sample and instrument requirements.

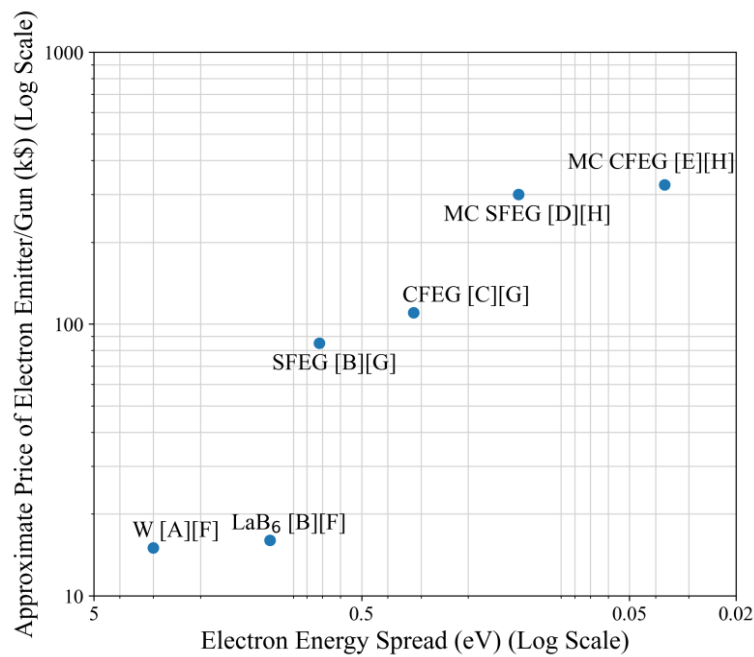

*Figure 2: A visual representation of the general trends in the price/performance trade-off of a variety of electron guns and monochromators. Further details for all emitters are found in the original literature [A](Carter & Williams, 2016), [B] (Stöger-Pollach, 2010), [C] (Sawada et al., 2009),[D] (Kisielowski et al., 2008), [E] (Carpenter et al., 2014). Pricing data was collected via personal communications; [F] R. Beanland, 7[th] December,2021. [G] G. Nicotra, November 29[th], 2021. [H] D. Muller, November 28[th], 2021.*

In order to give context to the financial aspect of this paper Figure 2 compares the electron energy spread of different electron guns and monochromated electron emitters (Carter &

Williams, 2016; Stöger-Pollach, 2010; Sawada et al., 2009; Kisielowski et al., 2008; Carpenter et al., 2014) versus their approximate cost which was gathered through private communication with the following researchers who have spent years working in the field of electron microscopy; R. Beanland, personal communications, 7[th] December, 2021; G. Nicotra, personal communications, November 29, 2021; D. Muller, personal communications, November 28, 2021. It is a visual representation of our literature findings and emphasises the differential cost between electron gun configurations.

From Figure 2 we see that cold FEGs have the lowest energy spread of all the electron sources but are the most expensive due to their ultra-sharp tips, more complicated set up, and requirement for a more complex pumping system (Franken et al., 2020). Electron monochromators are the most effective but also the most expensive solution to reducing electron energy spread and one must still consider the loss of electron dose when they are employed. These variations in price could be considered when comparing the simulated images produced by the different equipment.

## Materials and Methods

The energy spread of an electron source can be measured from the full width half maximum (FWHM) of the electron energy loss spectroscopy (EELS) zero loss peak (ZLP). For our analysis this was taken as 650 meV for a Schottky FEG (Grogger et al., 2008) and 287 meV for a cold FEG (Hachtel et al., 2018). Hachtel et al., 2018 demonstrated that as a monochromator reduces the electron energy spread, the FWHM of the ZLP decreases; this in turn decreases the integral of the ZLP, indicating a reduction of the total beam current. The beam current reduction of a Schottky and cold FEG from various levels of monochromation was determined from the EELS ZLP of a Schottky FEG (Figure 3 of (Grogger et al., 2008)) and a cold FEG (Figure 2a of (Hachtel et al., 2018)), respectively. The remaining beam current for both FEGs with increasing levels of monochromation was calculated by integrating the graph at their respective FWHM of various energy spreads (see Figure S1 in the supplemental). These values were then used in the following image simulations.

STEM image simulations were performed using the PRISM algorithm implemented in the Prismatic 2.0 package (DaCosta et al., 2021), an open source GPU-accelerated STEM simulation software. Across our simulations the appropriate wavelength was used corresponding to each acceleration voltage, as scattering cross section also depends on wavelength our simulations necessarily reflect changes in scattering with voltage. The simulation was averaged across 10 frozen phonons to account for thermal diffuse scattering. Using post-processing techniques chromatic aberration, finite source size, and Poisson noise were included in the simulated images. The chromatic aberration can be approximated by taking a weighted average of a spread of defocused images (Aarholt et al., 2020), and has been previously successfully executed using multislice (Sasaki et al., 2012). The defocus values and weightings were chosen from a Gaussian distribution (see Figure S2 in the supplementary information).

The annular dark-field (ADF) simulations used a detector angle range of 40-150 mrad. Of course, in simulations a range of any number of apertures could have been used. A practical microscope in the lab does not have a continuum of options, so a fixed-size aperture with a probe semi-angle of 28 mrad was chosen. The pixel dwell time was taken as 40 μs (Mullarkey et al., 2021), giving a total exposure time of ~8.35 seconds for each simulated image. Previously, Prismatic assumed a point source for electron emission at the gun, but in this work a source-size contribution was added. A mixed Gaussian-Cauchy distribution has been shown to accurately describe source effects (Verbeeck et al., 2012). However, as we are primarily interested in the trends *across* our data, a purely Gaussian distribution was chosen for simplicity where the source size (after demagnification) determines the FWHM of the Gaussian distribution. The source size for a Schottky FEG is lifetime dependant (Lebeau et al., 2009) and a mid-range estimate of 80 pm was chosen from literature (Lebeau et al., 2008, 2009; Brown et al., 2018; Erni et al., 2009). For the cold FEG a mid-range estimate of 40 pm was chosen (Sánchez-Santolino et al., 2018; Jones & Nellist, 2014; Brown et al., 2017). This source size contribution is added as a 2D convolution post-processing step to the ADF image. We are also aware that changing monochromator dispersion has some effect on effective source size however there is insufficient data on this in the literature so it was not modelled here. Finally, Poisson noise was added after scaling the relative dose by 1.5x in the case of the cold FEG due to its higher brightness (Xin et al., 2013; Carter & Williams, 2016).

Metallic nanoparticles are susceptible to knock-on damage (Azcárate et al., 2017). This makes them a very relevant candidate for optimising STEM performance at lower voltages. The simulations were of a gold nanoparticle on a carbon support imaged at three different low voltages (E = 15 keV, 30 keV, 60 keV). These acceleration voltages were arbitrarily chosen to span a wide range of low voltage imaging conditions, however the reader can adapt this value for their own particular sample. Here we considered only a $C_s$-corrected STEM where the $C_s$ coefficient was fully corrected (0 mm). The simulations for the Schottky FEG were evaluated for eleven different electron energy spreads spanning the range from severe monochromation to its usual electron energy spread ($\Delta E = 25$ meV, 50 meV, 75 meV, 110 meV, 150 meV, 200 meV, 250 meV, 287 meV, 400 meV, 500 meV, 650 meV), while the cold FEG was only evaluated for eight different energy spreads ($\Delta E = 25$ meV, 50 meV, 75 meV, 110 meV, 150 meV, 200 meV, 250 meV, 287 meV) as it had an initial lower energy spread of 287 meV. Our methodology does not aim to reproduce the energy spread of any particular monochromator, rather the energy spreads used in the simulation are spaced to reveal the underlying trends. The reader may wish to follow this methodology for their own particular energy spreads. These simulations were also run for four different objective lens $C_c$ coefficients to represent a variety of objective lenses available during tendering ($C_c = 1.1$ mm, 1.8 mm, and 3.0 mm (JEOL Ltd, 2004)) as well as, for comparison, a STEM with a chromatic aberration corrector installed ($C_c = 0$ mm) (see Figures S3 and S4 in the supplemental). As this is a simulation study, sample damage and scan distortion effects (Van Aert et al., 2019; Jones et al., 2015) aren't included, so the images produced will only show a best case scenario. However as this is consistent across all the data sets investigated it is still a fair comparison between all the simulations.

Since these were simulated images, the known ground truth was used to calculate the image signal-to-noise ratios (SNR) using

$$SNR = \frac{RMS(signal - mean(signal))}{RMS(noise - mean(noise))}, \quad (2)$$

where $RMS$ is the root mean square. The signal is defined as the noise-free ground truth image with source size but not Poisson noise or chromatic defocus blur, while the noise is defined as the difference once the signal is subtracted from the image inclusive of dose and chromatic effects. Both the signal and the noise are mean-subtracted to look at the undulations in the signal which give rise to the visual contrast. More details regarding the SNR calculation can be found in the supplemental information. These SNRs were plotted to determine the optimum energy spread for a STEM with either a Schottky or cold FEG based on its $C_c$ coefficient and the acceleration voltage. The same methodology we present here can be used by the reader to determine the optimum energy spread reduction for their STEM.

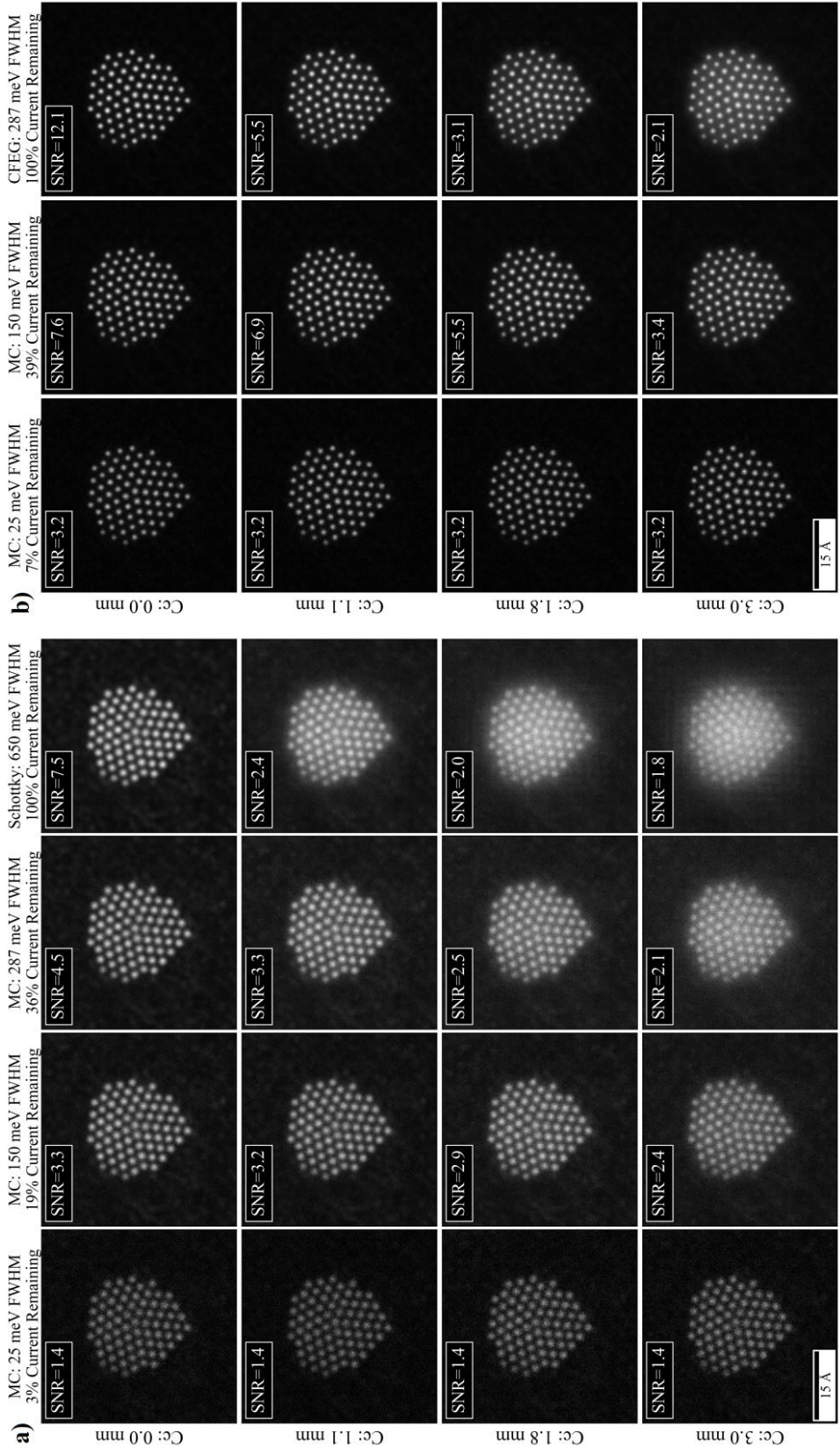

*Figure 3: Simulated Au nanoparticle images at (a) E = 30 keV with a Schottky FEG and (b) E = 60 keV with a cold FEG, both at various levels of monochromation. For each column ΔE = 25 meV, 150 meV, 287 meV, and an additional 650 meV column for the Schottky. The $C_c$ coefficient for each row is $C_c$ = 0 mm, 1.1 mm, 1.8 mm, and 3.0 mm from top to bottom respectively. Scale bar is 15 angstroms.*

## Results and Discussion

Figure 3a,b show a subset of our simulated gold nanoparticle imaged at 30 keV and 60 keV for a STEM with a Schottky and cold FEG respectively. Each row of images has been simulated with different objective lens $C_c$ coefficients and the columns show the simulations run at several of the different energy spreads of the monochromated Schottky and cold FEG. The $\Delta E = 650$ meV and $\Delta E = 287$ meV represent a sample imaged with the energy spread of an unmonochromated Schottky and cold FEG in Figure 3a and 3b respectively, while the $\Delta E = 287$ meV, 150 meV, and 25 meV in Figure 3a represent the sample imaged at three different levels of monochromation for the Schottky FEG. In Figure 3b only two levels of monochromation to $\Delta E = 150$ meV and 25 meV of the cold FEG are shown. For the full panel of all eleven and eight energy spreads at all the different acceleration voltages for the Schottky and cold FEG, please refer to Figures S3-S8 in the supplemental information. Comparing the different $\Delta E$ for the two FEGs at different levels of monochromation allows us to determine the optimum $\Delta E$ when imaging, while still considering the loss of beam current due to the monochromator.

One should note that the first three columns of the Schottky FEG in Figure 3a correspond to the same energy width for the corresponding columns of the cold FEG in Figure 3b. This highlights the two extremes of our simulations; the Schottky FEG with its lower brightness at a lower acceleration voltage of 30 kV and the cold FEG with its higher brightness at a higher acceleration voltage of 60 kV. Qualitatively examining Figure 3a we see that the energy spread of a Schottky FEG (650 meV) in a STEM with a high $C_c$ coefficient of 3.0 mm (bottom right corner) produces the lowest SNR out of the combinations. However, in the monochromated beam with the lowest energy spread (25 meV) and $C_c$ coefficient of 0.0 mm (top left corner), the beam current is attenuated, and the SNR suffers due to increased Poisson noise, meaning purely minimising both your energy spread and $C_c$ coefficient is not necessarily the optimal configuration. The same trend is seen for the 30 keV and 15 keV images in the supplemental except for the overall increased effect of chromatic aberration due to the reduced electron energy, resulting in reduced SNR. With the continual improvement in instrumentation, and an increased interest in Transmission SEM (Klein et al., 2012), as well as ultra-low voltage imaging in the STEM (Sasaki et al., 2014) analysis of voltages as low as 15 kV may be an interesting future research direction. A similar trend is seen for the cold FEG in Figure 3b and its images at lower acceleration voltages in Figures S7 and S8 in the supplemental.

To quantitatively analyse the simulations equation 2 was used to calculate the image SNRs as a function of energy spread for the four different $C_c$ coefficients in Figure 4 at 60 keV, 30 keV and 15 keV. Examining the right-hand side of the Schottky FEG $C_c = 3.0$ mm curve in Figure 4a we see that at the large energy spread of the Schottky FEG (650 meV), the SNR is low. As this energy spread, and consequently the effect of chromatic aberration for the monochromated Schottky FEG, decreases from right to left, the SNR increases until it reaches a maximum SNR. Conversely, as $\Delta E$ decreases, the beam current is further attenuated, reducing the number of electrons. Therefore for high levels of beam attenuation, this results in a lower SNR which decreases at a rate of the square root of the number of electrons due to the increasing effect of Poisson noise (Craven, 2011). The curves for the two lower Schottky FEG $C_c$ coefficients in Figure 4a follow the same trend except that their SNR peaks are shifted to higher electron energy spreads as their lower $C_c$ coefficients mean chromatic aberration will have less of an effect on their SNR. As a result, their maximum SNR is also higher. Naturally, when chromatic aberration has been corrected for ($C_c = 0$ mm), there is no reduction in SNR due to chromatic aberration as $\Delta E$ increases, so the SNR increases continuously as the dose is increased. The datapoints at $\Delta E = 650$ meV and $\Delta E = 287$ meV for the Schottky and cold FEG respectively may lie off the trendlines for all $C_c$ values as they are the unmonochromated and should be considered separately since any degree of monochromation will reduce the current heavily from the exclusion of the tails of the ZLPs.

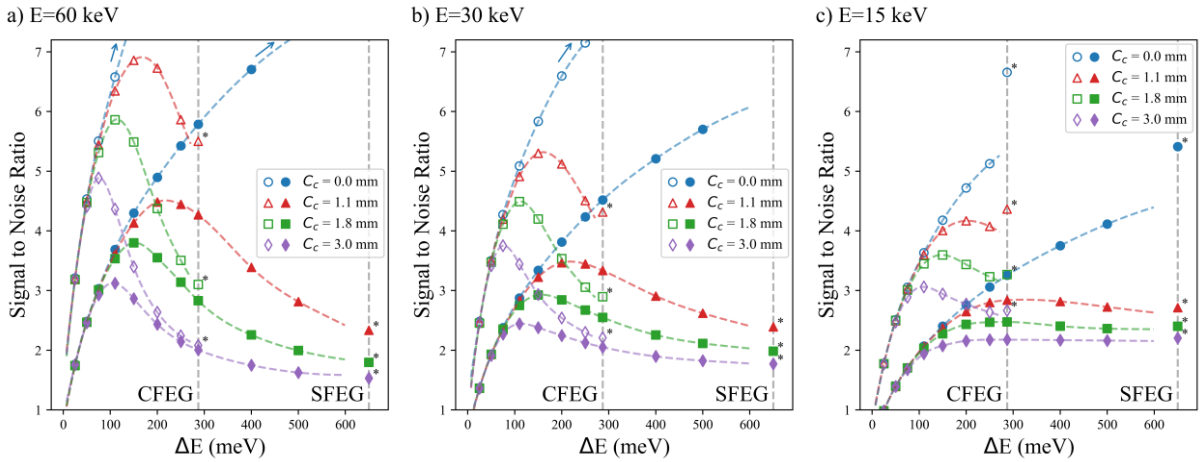

*Figure 3: The signal-to-noise ratio as a function of electron energy spread at (a) E = 60 keV, (b) 30 keV, and (c) E = 15 keV respectively. The filled in markers represent the Schottky FEG datapoints while the unfilled markers represent the cold FEG datapoints. Note that the datapoint marked with an asterisk lies off the trend as it's unmonochromated, whereas any form of monochromation will reduce the current heavily from the exclusion of the tails. This is not as visible for larger values of $C_c$ as these are dominated by chromatic effects.*

The monochromated cold FEG also follows the same trend as the monochromated Schottky FEG for different energy spreads and $C_c$ coefficients, however overall it has higher SNR values. This occurs because the cold FEG is brighter than a Schottky FEG (Xin et al., 2013; Carter & Williams, 2016) and as its initial energy spread is also inherently lower, monochromating both to an equivalent energy spread results in a lower beam current reduction for the cold FEG. The same trend in Figure 4a is followed for Figure 4b and c for the 30 keV and 15 keV images for both the Schottky and the cold FEG respectively, the only difference being that their SNRs are lower as they are imaged at a lower acceleration voltage and therefore chromatic aberration will have an even more deleterious effect on their SNR.

Figure 4 therefore reveals a clear optimum for each parameter set. In Figure 4a, for example, if your STEM had a Schottky FEG and a $C_c$ coefficient of 1.1 mm, the maximum SNR would be at a $\Delta E$ of approximately 190 meV, while it would be approximately 150 meV if your STEM had a cold FEG. If the $C_c$ coefficient was 3.0 mm however the $\Delta E$ would be required to be approximately 90 meV and 60 meV to achieve the maximum SNR for a Schottky and cold FEG respectively. As mentioned previously changing monochromator dispersion will have some effect on effective source size and this effect might cause the peak in this plot to shift by a small amount but the overall conclusion where we expect to find a peak is valid.

Analysing the trends across the non hardware $C_c$-corrected configurations, and where tilt range is not needed for the particular experiment, we find that a STEM with a high resolution objective lens with a small $C_c$ coefficient would be recommended. However, in a mixed-use facility which requires in-situ or tomography capabilities, equipping a pole piece with a very small gap may not be an option therefore installing a monochromator may be necessary and the associated condition of cost may be unavoidable. Anecdotally this is what we see in the community from the two dominant microscope manufacturers, with low voltage (S)TEM of metallic nanoparticles often being performed on Ultra High Resolution (UHR) pole-piece/cold FEG machines, while in situ experimentalists might prefer a wide gapped monochromated machines.

From Figure 4 it is also evident that the chromatic aberration corrected STEM ($C_c = 0$ mm) will produce the highest SNR in comparison to any uncorrected STEM due to the lack of effect of chromatic aberration. However as stated previously $C_c$ correctors are very expensive to install into the microscope and require extensive training for the STEM operator, these factors should therefore be carefully considered when deliberating their purchase.

It is also clear from Figure 4 that a monochromated cold FEG will always produce a higher SNR value than a monochromated Schottky FEG at the same $C_c$ coefficient and energy spread due to its higher brightness. However, it is interesting to note that the SNR of a low $C_c$ coefficient pole piece without monochromation at the energy spread of a cold FEG (287 meV) is still greater than that of a monochromated Schottky FEG STEM with a high $C_c$ coefficient pole piece. Therefore, if you have a high resolution pole piece with a small $C_c$ coefficient, and if an energy spread of 287 meV is acceptable, you do not necessarily have to invest in a monochromator to produce high SNR images at low voltages but could simply upgrade your Schottky FEG ($\Delta E = 650$ meV) to a cold FEG ($\Delta E = 287$ meV) to deliver the necessary performance.

The reader could apply this same method of analysing the performance of their microscope via the SNR of simulated images at various electron energy spreads for their own microscope's objective lens $C_c$ coefficient as well as choice of apertures and other instrument-dependent values. In this way the method followed here can be applied to more novel samples or the reader's own instrument, and an informed choice of what upgrade could be installed into their own particular machine to improve its performance could be achieved.

## Conclusion

Despite $C_s$ correctors now becoming increasingly commonplace in many modern facilities, low voltage imaging of sensitive materials remains a characterisation challenge. While determining the cost of any one component of a STEM is somewhat obfuscated, we find a strong correlation between the cost of a given illumination system and its energy spread performance. Here across four example hardware configurations and for three example acceleration voltages we show the expected signal-to-noise ratio as a function of varying degrees of monochromation for both a Schottky ($\Delta E \leq 650$ meV) and cold FEG ($\Delta E \leq 287$ meV). These plots show that for systems with a higher intrinsic $C_c$, more severe monochromation is indicated whilst for lower intrinsic $C_c$ a slightly less monochromated beam with more beam current yields a greater SNR. The plots also demonstrate that for a beam monochromated to any given energy spread, the cold FEG will always outperform the equivalent Schottky FEG due to increased brightness. Finally, for any given type of gun, a lower $C_c$ (smaller gap) pole piece will always give a higher SNR, but there may be other trade-offs accompanying the decision for example a decrease in EDX collection efficiency. With the escalating cost of flagship instrumentation, microscopists should be encouraged to look again at the various routes to achieve comparable levels of performance and with this methodology we offer a potential tool to do so. Revisiting concepts from decades past regarding pole piece geometry and emitter choice may be merited.

## Acknowledgments

The authors acknowledge the support of the Advanced Microscopy Laboratory of the Centre for Research on Adaptive Nanostructures and Nanodevices (CRANN). FQ is financially supported by the Provost's Project Award, PMB is supported by the School of Physics and the Advanced Materials and BioEngineering Research (AMBER) Centre (grant number 17/RC-PhD/3477), POD was supported by a School of Physics Summer Undergraduate Research Experience (SURE) scholarship. JJPP is supported by an SFI grant (grant number 19/FFP/6813) and LJ is supported by an SFI/Royal Society Fellowship (grant number URF/RI/191637).

# References

AARHOLT, T., FRODASON, Y. K. & PRYTZ, Ø. (2020). Imaging defect complexes in scanning transmission electron microscopy: Impact of depth, structural relaxation, and temperature investigated by simulations. *Ultramicroscopy* **209**, 112884.

VAN AERT, S., DE BACKER, A., JONES, L., MARTINEZ, G. T., BÉCHÉ, A. & NELLIST, P. D. (2019). Control of Knock-On Damage for 3D Atomic Scale Quantification of Nanostructures: Making Every Electron Count in Scanning Transmission Electron Microscopy. *Physical Review Letters* **122**, 066101.

AZCÁRATE, J. C., FONTICELLI, M. H. & ZELAYA, E. (2017). Radiation Damage Mechanisms of Monolayer-Protected Nanoparticles via TEM Analysis. *Journal of Physical Chemistry C* **121**, 26108–26116.

BROWN, H. G., CHEN, Z., WEYLAND, M., OPHUS, C., CISTON, J., ALLEN, L. J. & FINDLAY, S. D. (2018). Structure Retrieval at Atomic Resolution in the Presence of Multiple Scattering of the Electron Probe. *Physical Review Letters* **121**, 266102.

BROWN, H. G., ISHIKAWA, R., SÁNCHEZ-SANTOLINO, G., LUGG, N. R., IKUHARA, Y., ALLEN, L. J. & SHIBATA, N. (2017). A new method to detect and correct sample tilt in scanning transmission electron microscopy bright-field imaging. *Ultramicroscopy* **173**, 76–83.

CARPENTER, R. W., XIE, H., LEHNER, S., AOKI, T., MARDINLY, J., VAHIDI, M., NEWMAN, N. & PONCE, F. A. (2014). High energy and spatial resolution EELS band gap measurements using a nion monochromated cold field emission HERMES dedicated STEM. *Microscopy and Microanalysis* **20**, 70–71.

CARTER, C. B. & WILLIAMS, D. B. (2016). *Transmission electron microscopy: Diffraction, imaging, and spectrometry*. Springer.

CRAVEN, A. (2011). Details of STEM. In *Aberration-Corrected Analytical Transmission Electron Microscopy*, pp. 111–161. John Wiley & Sons, Ltd.

DACOSTA, L. R., BROWN, H. G., PELZ, P. M., RAKOWSKI, A., BARBER, N., O'DONOVAN, P., MCBEAN, P., JONES, L., CISTON, J., SCOTT, M. C. & OPHUS, C. (2021). Prismatic 2.0 – Simulation software for scanning and high resolution transmission electron microscopy (STEM and HRTEM). *Micron* **151**, 103141.

DREADEN, E. C., ALKILANY, A. M., HUANG, X., MURPHY, C. J. & EL-SAYED, M. A. (2012). The golden age: Gold nanoparticles for biomedicine. *Chemical Society Reviews* **41**, 2740–2779.

EGERTON, R. F. (2014). Choice of operating voltage for a transmission electron microscope. *Ultramicroscopy* **145**, 85–93.

EGERTON, R. F. (2019). Radiation damage to organic and inorganic specimens in the TEM. *Micron* **119**, 72–87.

ERNI, R., ROSSELL, M. D., KISIELOWSKI, C. & DAHMEN, U. (2009). Atomic-resolution imaging with a sub-50-pm electron probe. *Physical Review Letters* **102**, 096101.

FRANKEN, L. E., GRÜNEWALD, K., BOEKEMA, E. J. & STUART, M. C. A. (2020). A Technical Introduction to Transmission Electron Microscopy for Soft-Matter: Imaging,

Possibilities, Choices, and Technical Developments. *Small* **16**, 1906198.

GOODHEW, P. J., HUMPHREYS, J. & BEANLAND, R. (2000). *Electron Microscopy and Analysis*. CRC Press.

GROGGER, W., HOFER, F., KOTHLEITNER, G. & SCHAFFER, B. (2008). An introduction to high-resolution EELS in transmission electron microscopy. *Topics in Catalysis* **50**, 200–207.

HACHTEL, J. A., LUPINI, A. R. & IDROBO, J. C. (2018). Exploring the capabilities of monochromated electron energy loss spectroscopy in the infrared regime. *Scientific Reports* **8**, 1–10.

HAIDER, M., UHLEMANN, S., SCHWAN, E., ROSE, H., KABIUS, B. & URBAN, K. (1998). Electron microscopy image enhanced. *Nature* **392**, 768–769.

JAIN, P. K., HUANG, X., EL-SAYED, I. H. & EL-SAYED, M. A. (2008). Noble metals on the nanoscale: Optical and photothermal properties and some applications in imaging, sensing, biology, and medicine. *Accounts of Chemical Research* **41**, 1578–1586.

JEOL LTD (2004). JEM-2100 Instruction Manual. **1**, 179.

JONES, L. & NELLIST, P. D. (2014). Three-dimensional optical transfer functions in the aberration-corrected scanning transmission electron microscope. *Journal of Microscopy* **254**, 47–64.

JONES, L., YANG, H., PENNYCOOK, T. J., MARSHALL, M. S. J., VAN AERT, S., BROWNING, N. D., CASTELL, M. R. & NELLIST, P. D. (2015). Smart Align—a new tool for robust non-rigid registration of scanning microscope data. *Advanced Structural and Chemical Imaging* **1**, 1–16.

KIMOTO, K. & MATSUI, Y. (2002). Software techniques for EELS to realize about 0.3 eV energy resolution using 300 kV FEG-TEM. *Journal of Microscopy* **208**, 224–228.

KISIELOWSKI, C., FREITAG, B., BISCHOFF, M., VAN LIN, H., LAZAR, S., KNIPPELS, G., TIEMEIJER, P., VAN DER STAM, M., VON HARRACH, S., STEKELENBURG, M., HAIDER, M., UHLEMANN, S., MÜLLER, H., HARTEL, P., KABIUS, B., MILLER, D., PETROV, I., OLSON, E. A., DONCHEV, T., KENIK, E. A., LUPINI, A. R., BENTLEY, J., PENNYCOOK, S. J., ANDERSON, I. M., MINOR, A. M., SCHMID, A. K., DUDEN, T., RADMILOVIC, V., RAMASSE, Q. M., WATANABE, M., ERNI, R., STACH, E. A., DENES, P. & DAHMEN, U. (2008). Detection of single atoms and buried defects in three dimensions by aberration-corrected electron microscope with 0.5-Å information limit. *Microscopy and Microanalysis* **14**, 469–477.

KLEIN, T., BUHR, E. & GEORG FRASE, C. (2012). TSEM: A Review of Scanning Electron Microscopy in Transmission Mode and Its Applications. *Advances in Imaging and Electron Physics* **171**, 297–356.

KLIE, R. (2009). Reaching a new resolution standard with electron microscopy. *Physics* **2**, 85.

KRIVANEK, O. L., DELLBY, N., SPENCE, A. J., CAMPS, R. A. & BROWN, L. M. (1997). Aberration correction in the STEM. In *Electron Microscopy and Analysis Group Conference*, pp. 35–40. Cambridge.

KUNTSCHE, J., HORST, J. C. & BUNJES, H. (2011). Cryogenic transmission electron microscopy (cryo-TEM) for studying the morphology of colloidal drug delivery systems. *International Journal of Pharmaceutics* **417**, 120–137.

LEBEAU, J. M., FINDLAY, S. D., ALLEN, L. J. & STEMMER, S. (2008). Quantitative atomic resolution scanning transmission electron microscopy. *Physical Review Letters* **100**, 206101.

LEBEAU, J. M., FINDLAY, S. D., WANG, X., JACOBSON, A. J., ALLEN, L. J. & STEMMER, S. (2009). High-angle scattering of fast electrons from crystals containing heavy elements: Simulation and experiment. *Physical Review B - Condensed Matter and Materials Physics* **79**, 214110.

LINCK, M., HARTEL, P., UHLEMANN, S., KAHL, F., MÜLLER, H., ZACH, J., HAIDER, M., NIESTADT, M., BISCHOFF, M., BISKUPEK, J., LEE, Z., LEHNERT, T., BÖRRNERT, F., ROSE, H. & KAISER, U. (2016). Chromatic Aberration Correction for Atomic Resolution TEM Imaging from 20 to 80 kV. *Physical Review Letters* **117**, 076101.

LOUIS, C. & PLUCHERY, O. (2012). *Gold Nanoparticles for Physics, Chemistry and Biology*. Imperial College Press.

MULLARKEY, T., DOWNING, C. & JONES, L. (2021). Development of a Practicable Digital Pulse Read-Out for Dark-Field STEM. *Microscopy and Microanalysis* **27**, 99–108.

MULLER, D. A. & GRAZUL, J. (2001). Optimizing the environment for sub-0.2nm scanning transmission electron microscopy. *Journal of Electron Microscopy* **50**, 219–226.

MULLER, D. A., KIRKLAND, E. J., THOMAS, M. G., GRAZUL, J. L., FITTING, L. & WEYLAND, M. (2006). Room design for high-performance electron microscopy. *Ultramicroscopy* **106**, 1033–1040.

NAYDENOVA, K., MCMULLAN, G., PEET, M. J., LEE, Y., EDWARDS, P. C., CHEN, S., LEAHY, E., SCOTCHER, S., HENDERSON, R. & RUSSO, C. J. (2019). CryoEM at 100keV: A demonstration and prospects. *IUCrJ* **6**, 1086–1098.

SÁNCHEZ-SANTOLINO, G., LUGG, N. R., SEKI, T., ISHIKAWA, R., FINDLAY, S. D., KOHNO, Y., KANITANI, Y., TANAKA, S., TOMIYA, S., IKUHARA, Y. & SHIBATA, N. (2018). Probing the Internal Atomic Charge Density Distributions in Real Space. *ACS Nano* **12**, 8875–8881.

SASAKI, T., SAWADA, H., HOSOKAWA, F., SATO, Y. & SUENAGA, K. (2014). Aberration-corrected STEM/TEM imaging at 15kV. *Ultramicroscopy* **145**, 50–55.

SASAKI, T., SAWADA, H., OKUNISHI, E., HOSOKAWA, F., KANEYAMA, T., KONDO, Y., KIMOTO, K. & SUENAGA, K. (2012). Evaluation of probe size in STEM imaging at 30 and 60kV. *Micron* **43**, 551–556.

SAWADA, H., TANISHIRO, Y., OHASHI, N., TOMITA, T., HOSOKAWA, F., KANEYAMA, T., KONDO, Y. & TAKAYANAGI, K. (2009). STEM imaging of 47-pm-separated atomic columns by a spherical aberration-corrected electron microscope with a 300-kV cold field emission gun. *Journal of Electron Microscopy* **58**, 357–361.

SCHERZER, O. (1949). The theoretical resolution limit of the electron microscope. *Journal of Applied Physics* **20**, 20–29.

SCHRAMM, S. M., VAN DER MOLEN, S. J. & TROMP, R. M. (2012). Intrinsic instability of aberration-corrected electron microscopes. *Physical Review Letters* **109**, 163901.

SHARMA, P., BROWN, S., WALTER, G., SANTRA, S. & MOUDGIL, B. (2006). Nanoparticles for bioimaging. *Advances in Colloid and Interface Science* **123**–**126**, 471–485.

STÖGER-POLLACH, M. (2010). Low voltage TEM: Influences on electron energy loss spectrometry experiments. *Micron* **41**, 577–584.

TSUNO, K. & JEFFERSON, D. A. (1998). Design of an objective lens pole piece for a transmission electron microscope with a resolution less than 0.1 nm at 200 kV. *Ultramicroscopy* **72**, 31–39.

VERBEECK, J., BÉCHÉ, A. & VAN DEN BROEK, W. (2012). A holographic method to measure the source size broadening in STEM. *Ultramicroscopy* **120**, 35–40.

XIN, Y., KYNOCH, J., HAN, K., LIANG, Z., LEE, P. J., LARBALESTIER, D. C., SU, Y. F., NAGAHATA, K., AOKI, T. & LONGO, P. (2013). Facility implementation and comparative performance evaluation of probe-corrected TEM/STEM with schottky and cold field emission illumination. *Microscopy and Microanalysis* **19**, 487–495.

# Supplemental Information for Cost & Capability Compromises in STEM Instrumentation for Low-Voltage Imaging.

*Frances Quigley[1,2], Patrick McBean[1,2], Peter O'Donovan[1], Jonathan J. P. Peters[1,2], Lewys Jones[1,2]*

*1. School of Physics, Trinity College Dublin, Dublin 2, Ireland*

*2. Advanced Microscopy Laboratory, Centre for Research on Adaptive Nanostructures & Nanodevices (CRANN), Dublin 2, Ireland*

## Calculating the electron dose after the monochromation of the electron beam

The beam current reduction of a Schottky and cold FEG from various levels of monochromation was determined from the EELS ZLP of the Schottky and cold FEGs (Figure 3 of (Grogger et al. 2008) and Figure 3a of (Hachtel et al., 2018)) using the method detailed by (Hachtel et al., 2018). These can be seen on the left-hand plots of Figure S1 below. The remaining area was then integrated and compared to the total area to determine the fractional reduction in current, shown in the right-hand plot of Figure S1. These have been extended beyond the maximum 500 meV and 250 meV monochromation for the Schottky FEG and cold FEG, respectively, in order to display the full trend. For a given dwell time, the fractional reduction in current is equal to the fractional reduction in dose.

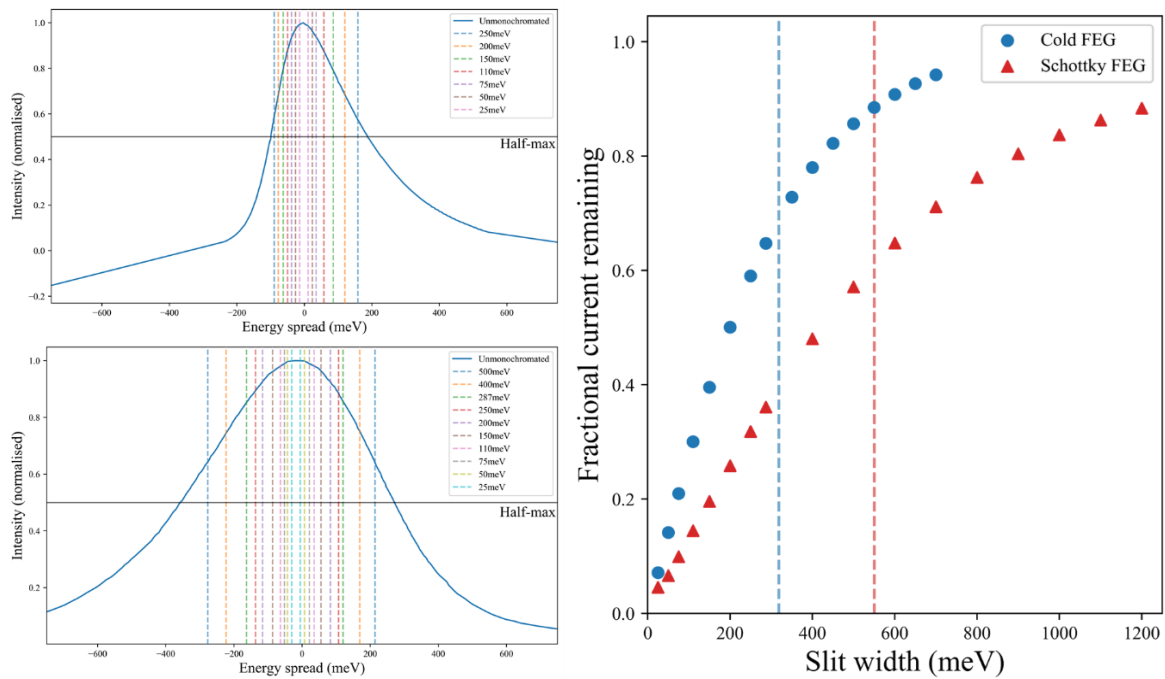

*Figure S1: (Left) ZLP of an unmonochromated (solid blue), and monochromated at various levels (dashed), of a cold FEG (top) and Schottky (bottom). The data points for the ZLP were taken from figure 3a in (Hachtel et al., 2018) and figure 3 in (Grogger et al., 2008) for the cold and Schottky FEG, respectively. (Right) By integrating the area after the different levels of monochromation, the relative current can be calculated and applied in terms of dose for Poisson noise.*

## Gaussian Distribution of weights

As detailed in (Aarholt et al., 2020), a series of defocus images were weighted according to a normalised Gaussian distribution and then summed. Figure S2 shows an example of this Gaussian distribution for one of the hardware configurations. 13 defocus planes were taken, evenly spanning a range of ±3σ (where σ is defined as $\sigma = \frac{FWHM}{2\sqrt{2\,ln(2)}}$ ).

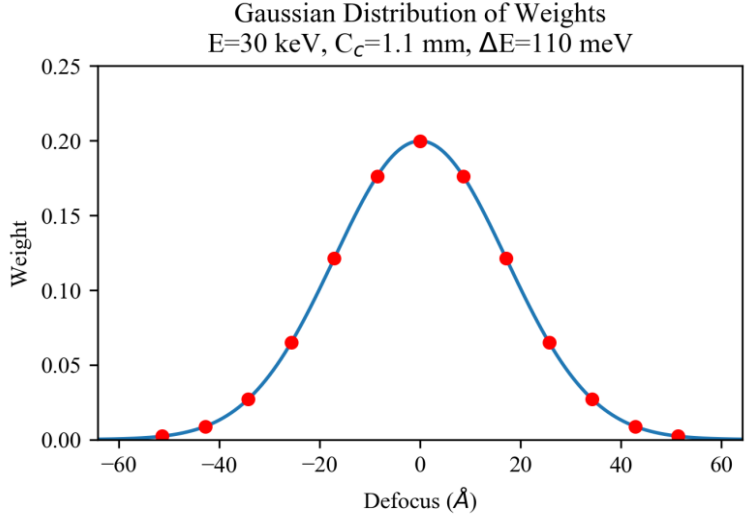

*Figure S2: Gaussian Distribution of weights for weighted summation of defocus planes. An equally spaced range of defocus values spanning a range of ±3σ. A normalised Gaussian is then used to assign weights to each of these defoci.*

## Signal-to-Noise Ratio Calculation

In order to calculate the SNR, the final image (with all effects included) and the ground truth image (i.e. the zero defocus plane image, with source size applied) are both mean-subtracted in order to examine only the undulations in the signal which produce the visual contrast. The ground truth was then subtracted from the final image to give just the noise. The SNR could then be calculated using the following equation;

$$SNR = \frac{RMS(signal - mean(signal))}{RMS(noise - mean(noise))}.$$

# Full Au Nanoparticle Images

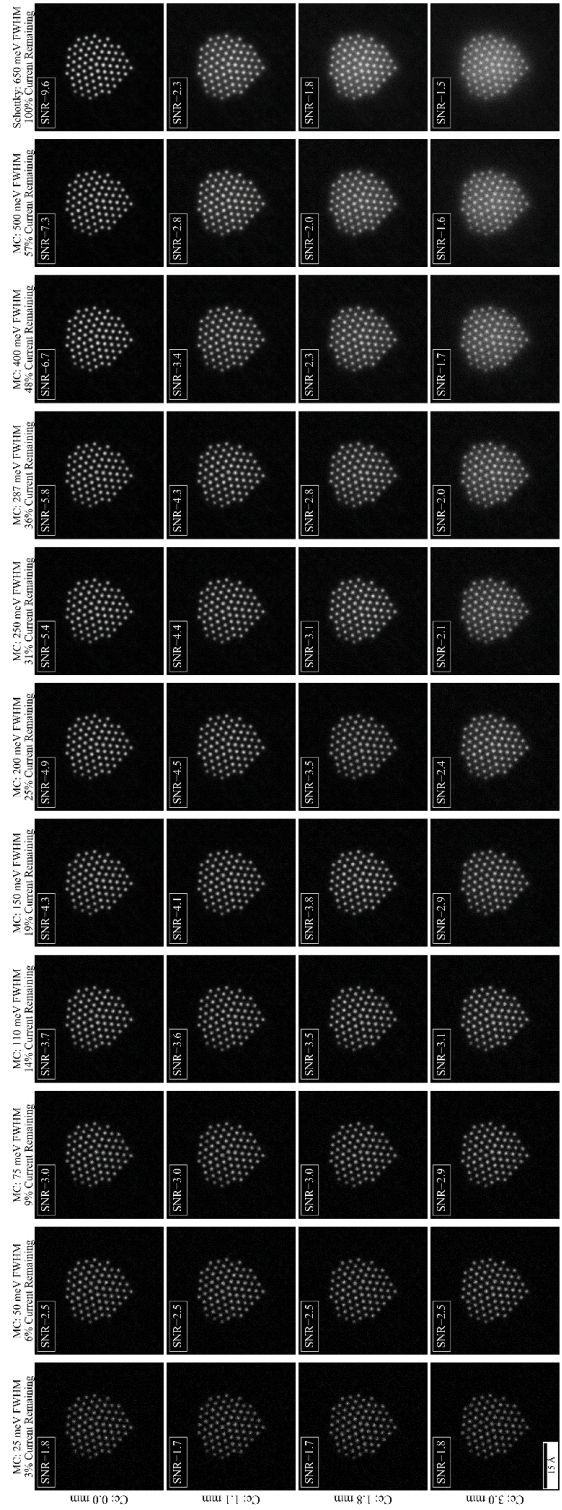

*Figure S3: Simulated Au nanoparticle images at E = 60 keV with a Schottky FEG at various levels of monochromation. For each column ΔE = 25 meV, 50 meV, 75 meV, 110 meV, 150 meV, 200 meV, 250 meV, 287 meV, 400 meV, 500 meV, and 650 meV. The C_c coefficient for each row is C_c = 0 mm, 1.1 mm, 1.8 mm, and 3.0 mm from top to bottom respectively. Scale bar is 15 angstroms.*

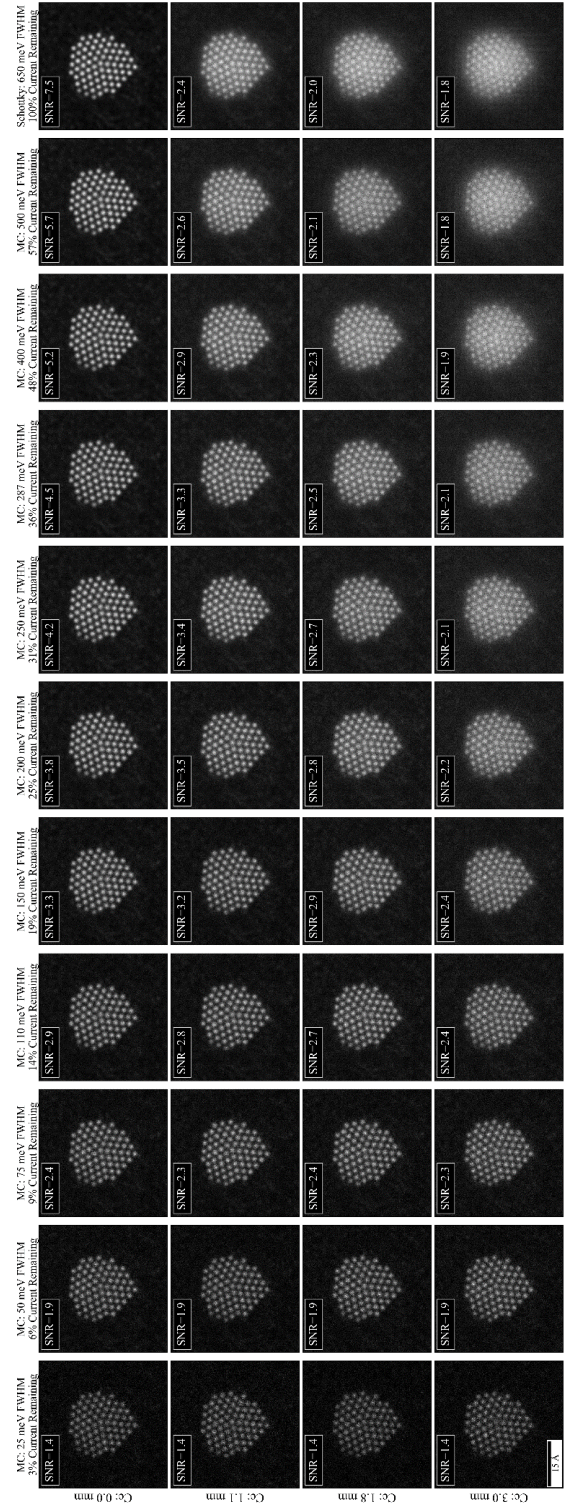

*Figure S4: Simulated Au nanoparticle images at E = 30 keV with a Schottky FEG at various levels of monochromation. For each column ΔE = 25 meV, 50 meV, 75 meV, 110 meV, 150 meV, 200 meV, 250 meV, 287 meV, 400 meV, 500 meV, and 650 meV. The $C_c$ coefficient for each row is $C_c$ = 0 mm, 1.1 mm, 1.8 mm, and 3.0 mm from top to bottom respectively. Scale bar is 15 angstroms.*

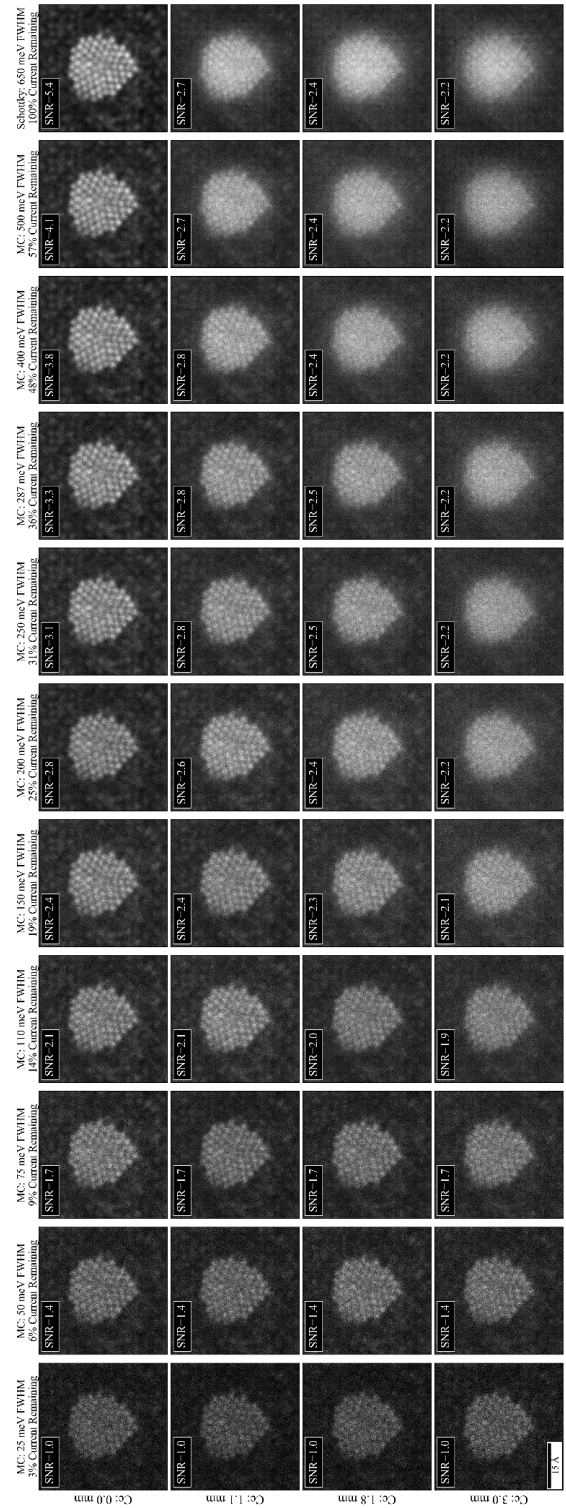

*Figure S5: Simulated Au nanoparticle images at E = 15 keV with a Schottky FEG at various levels of monochromation. For each column ΔE = 25 meV, 50 meV, 75 meV, 110 meV, 150 meV, 200 meV, 250 meV, 287 meV, 400 meV, 500 meV, and 650 meV. The $C_c$ coefficient for each row is $C_c$ = 0 mm, 1.1 mm, 1.8 mm, and 3.0 mm from top to bottom respectively. Scale bar is 15 angstroms.*

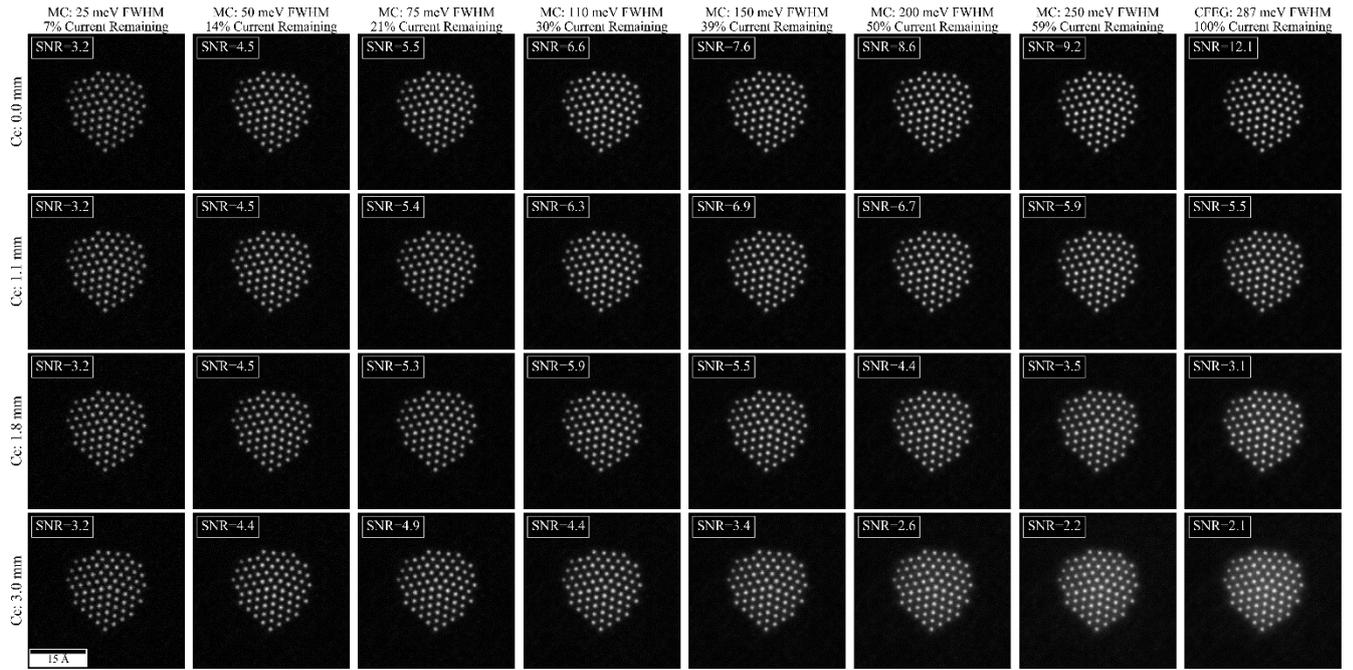

*Figure S6: Simulated Au nanoparticle images at E = 60 keV with a cold FEG at various levels of monochromation. For each column ΔE = 25 meV, 50 meV, 75 meV, 110 meV, 150 meV, 200 meV, 250 meV, 287 meV, 400 meV, 500 meV, and 650 meV. The $C_c$ coefficient for each row is $C_c$ = 0 mm, 1.1 mm, 1.8 mm, and 3.0 mm from top to bottom respectively. Scale bar is 15 angstroms.*

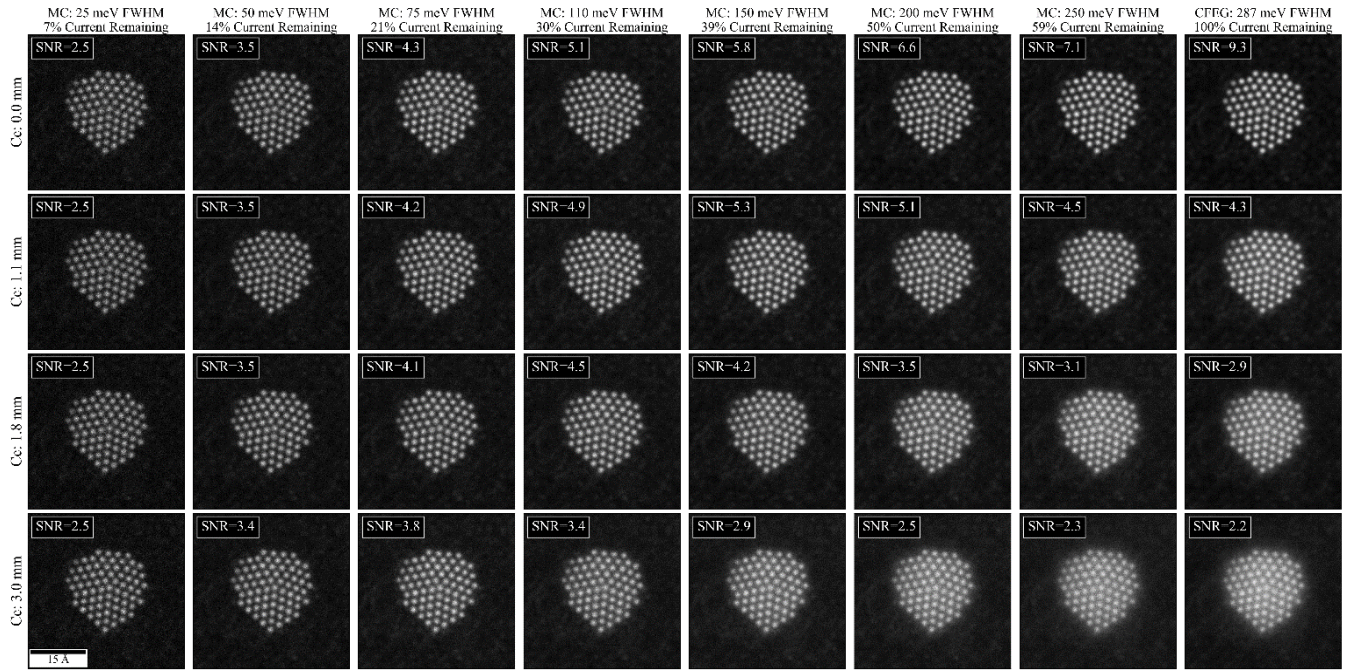

*Figure S7: Simulated Au nanoparticle images at E = 30 keV with a cold FEG at various levels of monochromation. For each column ΔE = 25 meV, 50 meV, 75 meV, 110 meV, 150 meV, 200 meV, 250 meV, 287 meV, 400 meV, 500 meV, and 650 meV. The $C_c$ coefficient for each row is $C_c$ = 0 mm, 1.1 mm, 1.8 mm, and 3.0 mm from top to bottom respectively. Scale bar is 15 angstroms.*

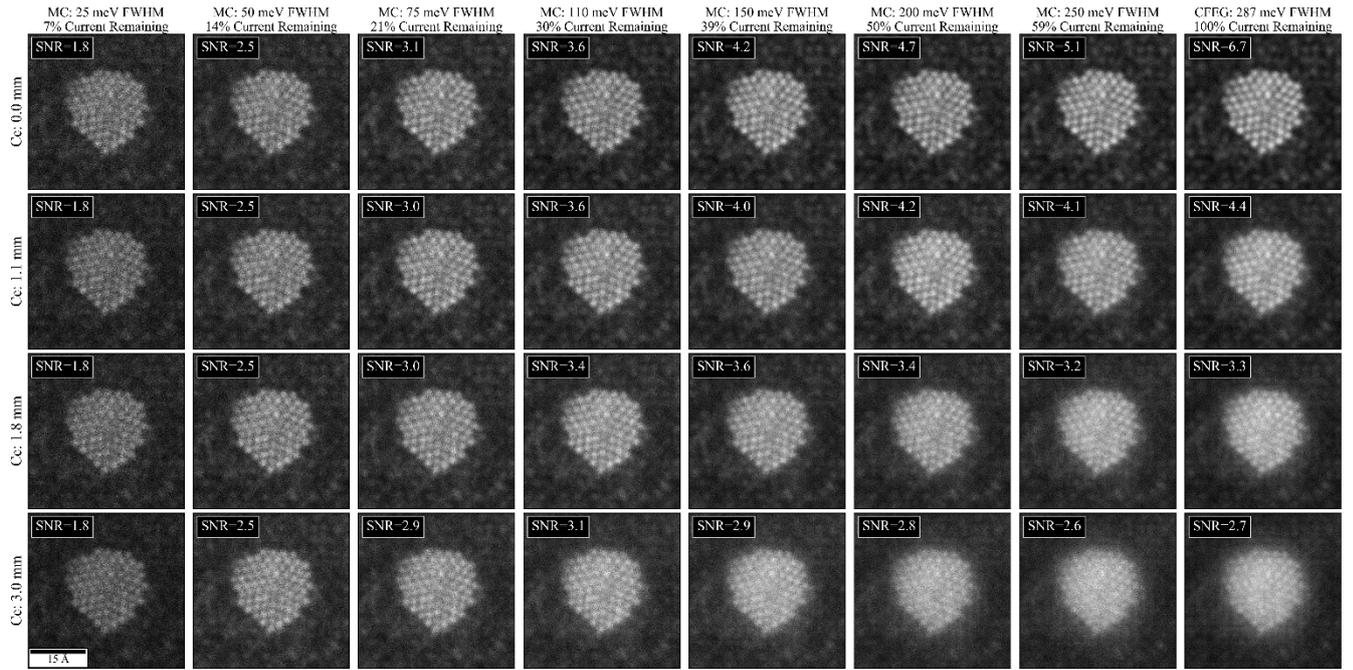

*Figure S8: Simulated Au nanoparticle images at E = 15 keV with a cold FEG at various levels of monochromation. For each column ΔE = 25 meV, 50 meV, 75 meV, 110 meV, 150 meV, 200 meV, 250 meV, 287 meV, 400 meV, 500 meV, and 650 meV. The $C_c$ coefficient for each row is $C_c$ = 0 mm, 1.1 mm, 1.8 mm, and 3.0 mm from top to bottom respectively. Scale bar is 15 angstroms.*